\documentclass[prb,twocolumn,preprintnumbers,amsmath,amssymb]{revtex4}

\usepackage{graphicx}
\usepackage{dcolumn}
\usepackage{bm}

\def\SRO{Sr$_2$RuO$_4$}
\def\RO{RuO$_2$}

\begin{document}

\preprint{Review of Scientific Instruments {\bf 75}, 1188 (2004)}

\title{Field-orientation dependent heat capacity measurements\\ at low temperatures with a vector magnet system}

\author{K. Deguchi}
\email{deguchi@scphys.kyoto-u.ac.jp}
\affiliation{Department of Physics, Graduate School of Science, Kyoto University, Kyoto 606-8502, Japan}

\author{T. Ishiguro}
\affiliation{Department of Physics, Graduate School of Science, Kyoto University, Kyoto 606-8502, Japan}

\author{Y. Maeno}
\affiliation{Department of Physics, Graduate School of Science, Kyoto University, Kyoto 606-8502, Japan}
\affiliation{Kyoto University International Innovation Center, Kyoto 606-8501, Japan}

\date{\today}

\begin{abstract}
We describe a heat capacity measurement system for the study of the field-orientation dependence for temperatures down to 50 mK. A ``vector magnet" combined with a mechanical rotator for the dewar enables the rotation of the magnetic field without mechanical heating in the cryostat by friction. High reproducibility of the field direction, as well as an angular resolution of better than $0.01^{\circ}$, is obtained. This system is applicable to other kinds of measurements which require a large sample space or an adiabatic sample environment, and can also be used with multiple refrigerator inserts interchangeably.
\end{abstract}


\maketitle

\section{INTRODUCTION}

The most important advances in superconductivity research over the past two decades have been the discoveries and studies of superconductors in which strong electronic correlations play crucial roles in both the normal and superconducting (SC) states. Thus, the so-called ``unconventional" superconductivity, realized in high-$T_{\rm c}$ cuprates, organics, heavy-fermion intermetallic compounds, ruthenate,\cite{Review} etc., has become one of the most actively studied topics of modern-day condensed matter physics. Since the anisotropic structure of the SC gap is directly connected with the SC order parameter and is closely related to the anisotropy of the pairing interaction, experimental determination of the gap structure is essential for the identification of the mechanism of superconductivity, especially among the unconventional superconductors. The field-orientation dependent specific heat, which is a direct measure of the quasiparticle density of states, is a powerful probe of the SC gap structure.\cite{Volovik,Vekhter}

	Such an experiment requires: (1) reaching a very low temperature compared with the SC transition temperature $T_{\rm c}$, (2) high angular resolution of exact field alignment with respect to crystallographic axes, and (3) high-resolution calorimetry in magnetic fields. To satisfy condition (1), we utilized a commercial dilution refrigerator. Unconventional superconductors often have low-dimensional electronic structures and exhibit strong anisotropy in the upper critical field $H_{\rm c2}$~(Ref.~\onlinecite{Tinkham}). Thus the angular resolution of better than $0.1^{\circ}$ is often required for the measurements. To control the field direction, the rotation of a sample with a geared goniometer has been widely used. A single-axis goniometer,\cite{NishiZaki,Dufeu} a low friction single-axis goniometer,\cite{Palm} and a double-axis goniometer\cite{lye,Settai,Herzog} have been used. As a single-axis goniometer with a very compact and simple structure, a piezoelectrically driven goniometer is also in use.\cite{Ohmichi} For rotation in liquid-helium mixtures, a miniature rotatable vacuum cell has been developed.\cite{Bondarenko} However, many of the geared goniometers cannot avoid the heat generated by mechanical friction and suffer from poor angular reproducibility caused by the backlash of the gears. To achieve the condition (2) and solve these problems, we developed a new system in which we electrically rotate the magnetic field with a ``vector magnet" and mechanically rotate this SC magnet with a one-circle goniometer against a fixed dilution refrigerator.

	The specific heat measurements in magnetic fields (3) have been carried out by a relaxation method\cite{Bachmann} with an automated small sample calorimeter\cite{Schwall,Griffing} from 50 mK to 8 K using a dilution refrigerator.\cite{Schultz,Steward} In the measurements, we often investigate the field amplitude $|\bm{H}|$ and field orientation $\phi, \theta$ dependence of the specific heat at constant temperatures and need to know the absolute value of the specific heat. In this article, we describe the construction, performance, and advantage of this system.

\section{DESIGN AND CONSTRUCTION}

\begin{figure}[t]
    \begin{center}
\includegraphics[width=8cm]{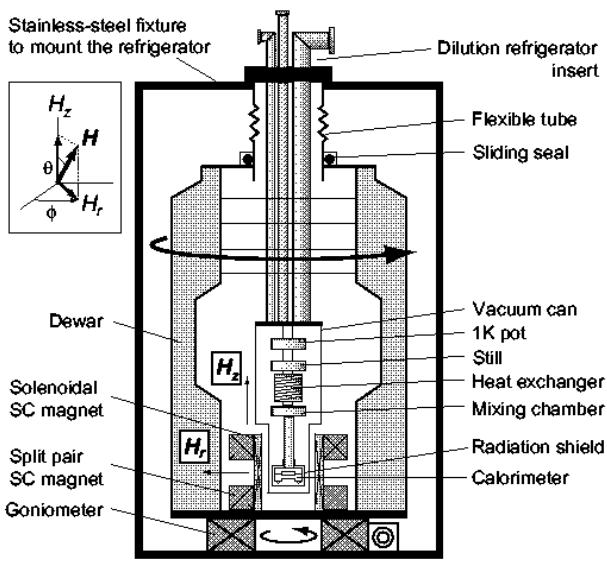}
    \end{center}
\caption{\label{fig:MAG} Schematic diagram of  the specific heat measurement system with a vector magnet.
}
\end{figure}

For controlling the field orientation at low temperature, we built a system with two orthogonally arranged SC magnets (Cryomagnetics, Inc., Oak Ridge TN, model VSC-3050) which generates the field $\bm{H}$ with a horizontal field $H_{r}$ and a vertical field $H_{z}$ to control the polar angle $\theta$ as shown in the inset of Fig.~\ref{fig:MAG}. A split pair and a solenoidal SC magnet provide $H_{r}$ of up to 5 T and $H_{z}$ of up to 3 T, respectively, with the homogeneity of $\pm$ 0.1\% over 10 mm diameter spherical volume. This vector magnet allows control of the field orientation $\theta = {\rm arctan}(H_{z}/H_{r})$ with a precision ${\mathit \Delta}\theta = 0.001^{\circ}$.
\begin{figure}[b]
    \begin{center}
\includegraphics[width=6cm]{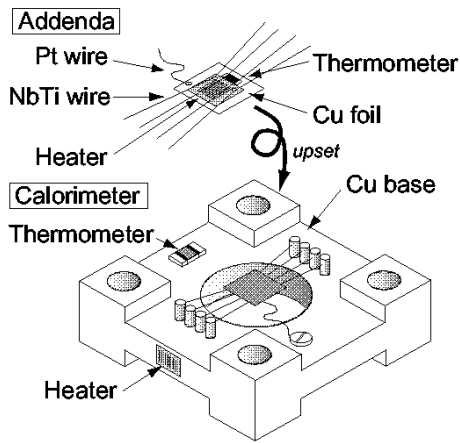}
    \end{center}
\caption{\label{fig:CELL} Schematic drawing of the calorimeter with an addenda and a copper base for the relaxation method.
}
\end{figure}

	The two SC magnets are installed at the bottom of a dewar seating on a goniometer (HUBER Diffraktionstecknik GmbH, Rimsting Germany, one-circle goniometer model 420) which controls the azimuthal angle $\phi$ of the field direction. The assembly can be rotated relative to a dilution refrigerator (Oxford Instruments, Inc., Witney, Oxon, UK Kelvinox 25) which is fixed, with a minimum step of ${\mathit \Delta}\phi = 0.001^{\circ}$. The goniometer can rotate the two SC magnets and the dewar with the magnetic field applied either in a persistent-current mode or in a current feeding mode. We used long enough magnet cables to allow rotation of $\pm 360^{\circ}$. Controlling the two magnets and the goniometer, we can rotate and align the field $\bm{H}$ continuously to any direction. For example, for the study of a system with a two-dimensional (2D) electronic structure, we can rotate the field continuously within the 2D plane with a misalignment no greater than $0.01^{\circ}$ from the plane. Advantages of the field rotating system with the vector magnet and the goniometer, compared with a sample rotating system, are: (1) A large sample space, (2) applicability to different refrigerator inserts interchangeably, (3) no mechanical heating by friction at low temperatures except for the eddy current heating by variations of the magnetic field, and (4) good reproducibility of the field orientation.

	Details of the addenda of the calorimeter are schematically shown in Fig.~\ref{fig:CELL}. The thermometer is made of a 1.8 k$\Omega$ \RO\ thick-film resistor (KOA, RK73K1EJ).\cite{KOA} To maximize the thermal response, as well as to minimize the heat capacity of addenda, the alumina substrate of the \RO\ thick-film resistor is polished and thinned down to the size of $1.0 \times 0.5 \times 0.1 \;{\rm mm}^{3}$. The solder on the terminal of the \RO\ thick-film resistor is also shaved to eliminate a change of addenda heat capacity at the SC transition of the solder. The heater is made of a 350 $\Omega$ strain gauge (Kyowa, KFG-2-350-C1-16N2C2)\cite{KYOWA} which is cut out to the size $2.5 \times 2.5 \times 0.05 \;{\rm mm}^{3}$. The thermometer and heater are glued directly on one side of a thin foil of oxygen-free high-conductivity copper with 5N purity, $3.0 \times 3.0 \times 0.025 \;{\rm mm}^{3}$, using a minimal amount of epoxy (Emerson and Cuming, Stycast 1266). The sample is placed on the other side of the copper foil with a small amount of grease (Apiezon Products, N grease). The electrical leads are monofilament niobium-titanium wires, 67 $\mu$m in diameter, with a SC transition temperature $T_{\rm c} = 9.3$ K and upper critical field $\mu_{0}H_{\rm c2} = 12$ T. We use this wire in the SC state in order to eliminate the parasitic heating as well as to suppress the heat leak. These wires are attached to the thermometer and heater by heat-cure-type silver epoxy (Dupont, 6838). As shown in Fig.~\ref{fig:CELL}, a platinum wire with the diameter of 10 $\mu $m is connected with normal-cure-type silver epoxy (Dupont, 4922N) as a thermal link between the addenda and the copper base, which serves as a thermal bath. The calorimeter is covered with a copper foil as a radiation shield.

	Heat capacity measurements are carried out with a relaxation method\cite{Bachmann} as follows. When we apply a constant power $P_0$ to the addenda, the temperature increases from the base temperature $T_0$ to $T_0 + {\mathit \Delta}T$ at a steady state. The thermal conductance $K$ of the wires is evaluated as $K=P_0/{\mathit \Delta}T$. By switching off the power, the temperature relaxes from $T_0 + {\mathit \Delta}T$ down to $T_0$ according to
\begin{eqnarray}
T(t) = T_0 + {\mathit \Delta}T{\rm exp}\;(-t/\tau _1).
\end{eqnarray}
We determine the relaxation time $\tau _1$ from the fitting to a relaxation curve in the range of $0<t<10\tau _1$. We typically obtain $\tau _1 \simeq 25$ s in the heat capacity measurement of aluminum and Sr$_2$RuO$_4$ as described below. Thus, the total heat capacity can be calculated from $C = K\tau_1$. The simple model given above yields good results in most cases, but if the thermal conductance between the sample and addenda cannot be made sufficiently large compared with that of the thermal link, the relaxation curve is characterized by a large initial slope compared to the rest of the decay, which is called the $\tau _2$ effect.\cite{Anderson,Shepherd} In that case, the relaxation curve is better described by
\begin{eqnarray}
T(t) = T_0 + A\;{\rm exp}(-t/\tau _1) + B\;{\rm exp}(-t/\tau _2).
\end{eqnarray}
The total heat capacity is described as $C = K(A\tau_1 + B\tau_2)/{\mathit \Delta}T$ with $A + B = {\mathit \Delta}T$. Improving the thermal contact between the assembly of addenda and a sample results in very small $\tau _2$ effect: We typically obtain $\tau _2/\tau _1 < 0.05$ and $B/A < 0.05$ in the range of 50 mK $<T<$ 8 K. Thus the total heat capacity is approximated as $C \simeq K\tau_1\frac{A}{{\mathit \Delta}T}$. Since precise measurements of short $\tau_2$ are difficult with nonlinear fits of relaxation curves and $\frac{A}{{\mathit \Delta}T} \approx 1$ is obtained in our case, we use this approximated equation. Therefore, the accuracy of heat capacity measurement with a relaxation method is well assured in this apparatus.
\begin{figure}[t]
    \begin{center}
\includegraphics[width=7cm,clip]{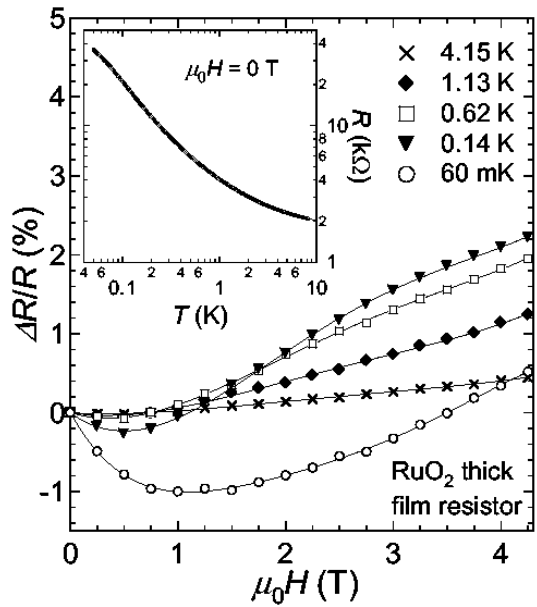}
    \end{center}
\caption{\label{fig:RuO2H} Magnetoresistance ${\mathit\Delta} R/R=[R(H,T)-R(0,T)]/R(0,T)$ vs magnetic field for a \RO\ thick film resistor at several temperatures. Inset: Temperature dependence of resistance of a \RO\ thick film resistor in zero field.  
}
\end{figure}

	For a relaxation curve measurement, which requires fast readings of the thermometer resistance, we use a phase-sensitive detector (Stanford Research, model SR830)  with the measurement frequency of 439 Hz and a 30 ms time constant. To give a heat pulse and measure the heater power, a dc current source (Yokogawa, model 7651) and a nano-voltmeter (Keithley, model 2182) are connected to the addenda heater.  In order to accurately measure the specific heat, it is very important to maintain a constant temperature of the thermal bath. For this purpose, the temperature of the copper base is measured with a thermometer (a 1.8 k$\Omega$ \RO\ thick film resistor) and an ac resistance bridge (RV-Elektroniikka, model AVS-47) using the excitation frequency of 15 Hz, while it is controlled with a heater (strain gauges with 350 $\Omega$) and a temperature controller (RV-Elektroniikka, model TS-530). To avoid noise pick up and rf heating of the sensors, the signal wires are filtered at the top of the cryostat at room temperature by means of simple low-pass $LC$ (inductance-capacitance) filters with a cut off frequency of 10 kHz. Consequently, we achieved the thermometry with low noise and high stability at low temperatures.
\begin{figure}[t]
    \begin{center}
\includegraphics[width=7cm,clip]{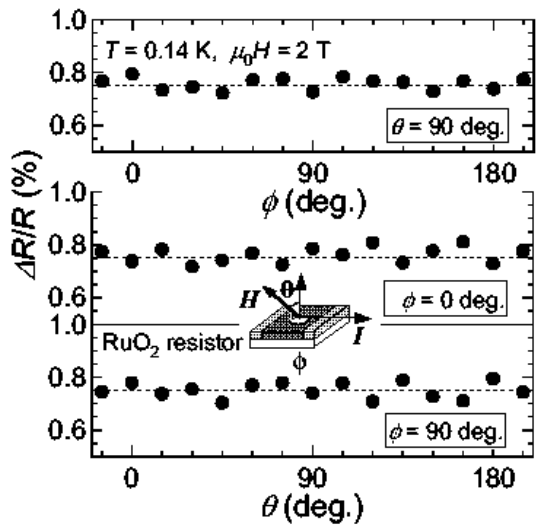}
    \end{center}
\caption{\label{fig:RuO2ANG} Magnetoresistance ${\mathit\Delta} R/R=[R(H,T)-R(0,T)]/R(0,T)$ vs magnetic field orientation for a \RO\ thick-film resistor at $T = 0.14$ K and $\mu_0H = 2$ T. The polar angle $\theta$ is defined as the angle between the field direction and the axis perpendicular to the \RO\ film plane, while the azimuthal angle $\phi$ is defined as the angle from the current direction. 
}
\end{figure}

\section{CALIBRATION}

The calorimetry in magnetic field requires appropriate calibrations of the resistance thermometer for its magnetoresistance ${\mathit\Delta} R/R=[R(H,T)-R(0,T)]/R(0,T)$. In addition, the examination of {\it anisotropy} of its magnetoresistance is important in the study of the field-orientation dependence of the specific heat. The \RO\ thick-film resistor with a nominal resistance of 1.8 k$\Omega$ (KOA, RK73K1EJ) was adopted as the resistance thermometer of the calorimetry in magnetic field because of its advantages of high sensitivity at low temperature, small heat capacity, fast thermal response time, and low cost.\cite{Doi,Bosch,Li,Willekers,Uhlig,Goodrich,Zhang,Fortune} For the calibration of the thermometer in zero field, commercially calibrated \RO\ thermometer (Scientific Instruments, model RO600A2) was used in the range from 50 mK to 8 K. The inset in Fig.~\ref{fig:RuO2H} shows the temperature dependence of resistance of the addenda thermometer in zero field. It is well represented by a characteristic temperature dependence of the electrical conduction mechanism due to variable range hopping in three-dimensional Anderson localized states\cite{Localized}: $R=R_0\;{\rm exp}(T_0/T)^{1/4}$.

	Field-strength and field-orientation dependences of the \RO\ thick film resistor were measured at a constant temperature. To maintain a constant temperature, each measurement was carried out in the persistent-current mode of the SC magnet after very slow field sweeping and establishing a thermal equilibrium state. Figure~\ref{fig:RuO2H} shows the field-strength dependence of the magnetoresistance at several temperatures. At low temperatures, the magnetoresistance is negative at low fields but becomes positive at higher fields, consistent with previous reports.\cite{Li,Uhlig,Goodrich,Zhang} Such behavior is attributable to disorder effects (weak localization) at low temperatures, which produce negative magnetoresistance in the low-field region.\cite{AL,Lee} At high fields, the field dependence is nearly proportional to the square root of $H$, which originates from weak localization with electron-electron scattering.\cite{Goodrich}
\begin{figure}[t]
    \begin{center}
\includegraphics[width=7cm,clip]{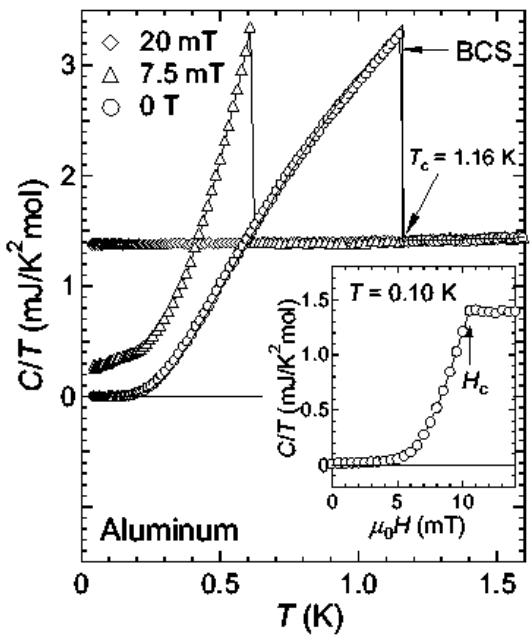}
    \end{center}
\caption{\label{fig:ALHT}Temperature dependence of the specific heat divided by temperature $C/T$ of aluminum in several magnetic fields. Inset: Field dependence of $C/T$ at $T = 0.10$ K.
}
\end{figure}

	Figure~\ref{fig:RuO2ANG} shows the anisotropy of the magnetoresistance at 0.14 K, investigated in three ways:  The azimuthal angle $\phi$ dependence within the plane of the \RO\ film, the polar angle $\theta$ dependence from the current direction, and the polar angle $\theta$ dependence perpendicular to the current direction. As is clear from Fig.~\ref{fig:RuO2ANG}, no significant field-orientation dependence against the \RO\ chip and current directions was found. This feature provides an excellent advantage to the thermometry in magnetic fields. The information in Fig.~\ref{fig:RuO2ANG} explicitly represents full sets of data on the field-orientation dependence of a \RO\ for the first time, which is consistent with comments in previous reports.\cite{Doi,Willekers} Since the dependence of $\theta$ and $\phi$ can be neglected, the resistance $R(T,H)$ is expressed in terms of a 2D-polynomial function of  temperature $T$ and field-strength $H$~(Refs.~\onlinecite{Zhang,Fortune}): 
\begin{eqnarray}
\displaystyle{\rm ln}\;R=\sum_{m=0}^{6}\sum_{n=0}^{4}a_{m,n}H{^m}({\rm ln}\;T)^n.
\end{eqnarray}
The coefficients $a_{m,n}$ are determined from the data taken from the field sweeps at 12 temperatures. A part of the results of fitting made in the range of 50 mK $<T<$ 8 K, 0 T $<\mu_0H<$ 4.25 T are shown in Fig.~\ref{fig:RuO2H}.
\begin{table}[t]
\caption{\label{tab:table1}Comparison of normal-state and SC-state parameters of aluminum with previous report.}
\begin{ruledtabular}
\begin{tabular}{clll}
&&This work&Phillips\cite{Phillips}\\
\hline
$\gamma$&$({\rm mJ/K^2mol})$&1.37&1.35\\
${\it \Theta}_{\rm D}$&$({\rm K})$&420&427.7\\
$T_{\rm c}$&$({\rm K})$&1.16&1.163\\
${\mathit \Delta}C/{\it \gamma}T_{\rm c}$&&1.37&1.32\\
$H_{\rm c}$&$({\rm mT})$&10.5&10.3\\
\end{tabular}
\end{ruledtabular}
\end{table}
\begin{figure}[b]
    \begin{center}
\includegraphics[width=7.5cm,clip]{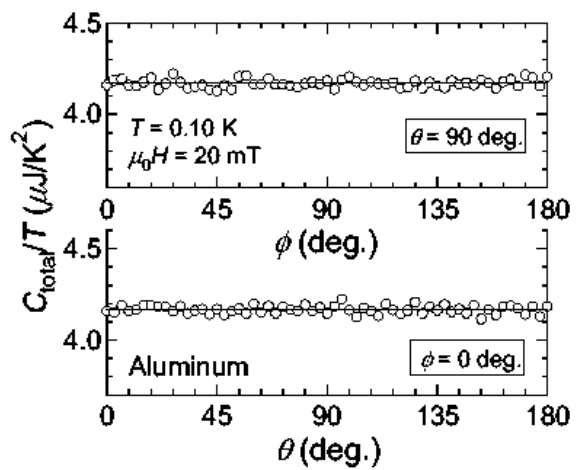}
    \end{center}
\caption{\label{fig:ALANG} Field-orientation dependence of the total specific heat $C_{\rm total}=C+C_{\rm addenda}$ divided by temperature of the aluminum in the normal state at $T = 0.10$ K and $\mu_0H = 20$ mT.
}
\end{figure}

\section{PERFORMANCE}

In order to demonstrate its performance, we examine the accuracy of the measured specific heat of aluminum (Al), a conventional Type-I superconductor with $T_{\rm c}=1.17$ K. We also inspect the field-orientation dependence of the heat capacity of the addenda. A polycrystalline aluminum sample with a purity of 6N was cut from the polycrystalline rod, to a size of $3.9 \times 3.9 \times 1.9 \;{\rm mm}^{3}$ with 77.9 mg. The specific heat $C$ of the sample was evaluated from the total heat capacity $C_{\rm total}$ after subtraction of the addenda heat capacity $C_{\rm addenda}$. The heat capacity of the addenda was separately measured and is described as $C_{\rm addenda} = (\alpha +\alpha^\prime H^2)/T^2 + \gamma T + \beta T^3$ with $\alpha=1.03 \times 10^{-4}\;\mu{\rm JK}$, $\alpha^\prime=4.68 \times 10^{-5}\;\mu{\rm JK/T^2}$, $\gamma=1.15 \times 10^{-1}\;\mu{\rm J/K^2}$, and $\beta=4.36 \times 10^{-2}\;\mu{\rm J/K^4}$.

	Figure~\ref{fig:ALHT} shows the temperature dependence of $C/T$ of the aluminum sample in magnetic fields. At $\mu_0H =$ 0 T and 7.5 mT, the SC transitions were observed at $T_{\rm c}=$ 1.16 K and 0.62 K, respectively. At $\mu_0H =$ 20 mT, the normal state persists down to 50 mK. From the specific heat in the normal state, we can estimate the electronic specific heat coefficient $\gamma=1.37\;{\rm mJ/K^2mol}$ and the Debye temperature ${\it \Theta}_{\rm D} = 420$ K from the fitting to $C = \gamma T + \beta T^3,  \beta = (12/5)\pi ^4 R{\it \Theta}_{\rm D}^{-3}$ below 4 K. In zero field, the specific heat jump ${\mathit \Delta}C/{\it \gamma}T_{\rm c} = 1.37$ was observed with a SC transition at $T_{\rm c}=1.16$ K. These results are in excellent agreement with a previous report by an adiabatic method using a sample weighting 212.64 g,\cite{Phillips} as summarized in Table~\ref{tab:table1}. The temperature dependence of the specific heat is fully consistent with theoretical calculation based on the Barden-Cooper-Schrieffer theory\cite{Muhlschlegel} in all temperature range. This examination assures the high accuracy and high resolution in the specific heat measurement with this system. The requirement of the entropy balance suggests that the transition under the magnetic field ($\mu_{0}H = 7.5$ mT) is of the first order with a latent heat of $L = T_{\rm c}{\mathit \Delta} S = 0.091$ mJ/mol.

	The inset of Fig.~\ref{fig:ALHT} shows the field dependence of $C/T$ at $T = 0.10$ K. $C/T$ is nearly equal to the electronic contribution $C_{\rm e}/T \propto$ density of state (DOS) at low temperatures. Since a fully gapped SC gap opens and removes the DOS at the Fermi surface, $C/T$ remains nearly zero at low fields. With an increasing magnetic field, the DOS increases because of the normal state region in the intermediate state, and $C_{\rm e}/T \propto$ DOS recovers up to $\gamma$ at the critical field $H_{\rm c}=10.5$ mT. This value is also consistent with the previous report\cite{Phillips} and guarantees the accuracy of the applied magnetic field amplitude generating by the vector magnet.

	Figure~\ref{fig:ALANG} shows the field-orientation dependence of the total specific heat divided by temperature $C_{\rm total}/T$ with the aluminum sample in the normal state at $T = 0.10$ K and $\mu_0H = 20$ mT. The fraction of the addenda heat capacity is $C_{\rm addenda}/C_{\rm total} = 0.08$. The definitions of  $\theta$ and $\phi$ are with respect to the addenda thermometer as shown in Fig.~\ref{fig:RuO2ANG}. Since the polycrystalline aluminum in the normal state does not give the field-orientation dependence of the total heat capacity, these results reveal that the addenda heat capacity has totally negligible field-orientation dependence.

\section{APPLICATION EXAMPLE}

As an application example, we describe an experiment for determination of the SC gap structure of \SRO\ from the field-orientation dependence of the specific heat. The layered perovskite \SRO\ is an unconventional Type-II superconductor with strong quasi-two dimensionality.\cite{Maeno1} Its superconductivity exhibits pronounced unconventional features attributable to spin-triplet {\it p}-wave pairing with time-reversal symmetry breaking, or ``chirality", associated with the Cooper-pair orbital moment.\cite{Review} In spite of a large number of experiments and theories on its SC gap structure and spin-triplet superconductivity, the details of the gap structure still remain to be clarified.

	Single crystals of \SRO\ were grown by a floating-zone method in an infrared image furnace.\cite{growth} The sample with $T_{\rm c} = 1.48$ K, close to the estimated value for impurity and defect-free specimen ($T_{\rm c0}=1.50$ K), was chosen for detailed study. This crystal was cut and cleaved from a single crystalline rod, to a size of $2.8 \times 4.8 \;{\rm mm}^{2}$ in the $ab$ plane and $0.50 \;{\rm mm}$ along the $c$ axis, and weighs 39.8 mg. An x-ray rocking curve of the sample shows the characteristics of a single crystal of high quality; the diffraction peak width [full width at half maximum (FWHM)] being comparable to that for a Si crystal (with a FWHM of $0.06 ^{\circ}$) limited by the instrumental resolution of our diffractometer. The directions of the tetragonal crystallographic axes of the sample were determined by x-ray Laue pictures. The side of the crystal was intentionally misaligned from the [110] axis by $16^{\circ}$ to distinguish the essential anisotropy related to the unconventional superconductivity from the extrinsic anisotropy due to surface effects.
\begin{figure}[t]
    \begin{center}
\includegraphics[width=8cm,clip]{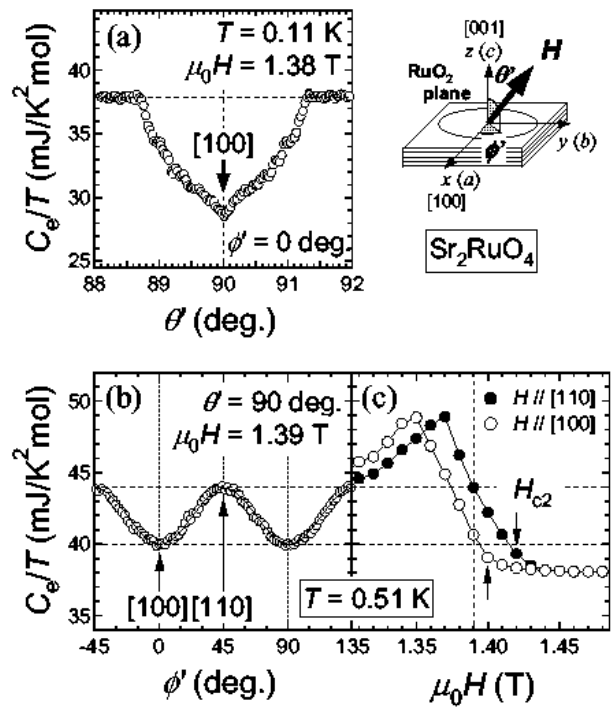}
    \end{center}
\caption{\label{fig:ALIGN}(a) The polar angle $\theta^\prime$ dependence of the electronic specific heat divided by temperature $C_{\rm e}/T$ of  Sr$_2$RuO$_4$. (b) The azimuthal angle $\phi^\prime$ dependence of $C_{\rm e}/T$. (c) Field dependence of $C_{\rm e}/T$ in the vicinity of the upper critical field $H_{\rm c2}$.
}
\end{figure}

	In this experiment, capability of rotating $\bm{H}$ within the RuO$_2$ plane with high accuracy is very important because of strong quasi-2D electronic structure with cylindrical Fermi surfaces. A slight field misalignment causes extrinsic two-fold anisotropy of the field-orientation dependent specific heat due to a large $H_{\rm c2}$ anisotropy ($H_{{\rm c2}\parallel ab}/H_{{\rm c2}\parallel c}\approx 20$).\cite{Deguchi,Yaguchi} For specifying the direction of the applied magnetic field with respect to the crystallographic axes, we shall introduce the polar angle $\theta^\prime$, for which $\theta^\prime = 0^{\circ}$ corresponds to the [001] direction, and the azimuthal angle $\phi^\prime$, for which $\phi^\prime = 0^{\circ}$ with $\theta^\prime = 90^{\circ}$ corresponds to the [100] direction. It is noted that the tetragonal symmetry of the crystal structure is conserved down to temperatures as low as 110 mK.\cite{Gardner}
\begin{figure}[t]
    \begin{center}
\includegraphics[width=8cm,clip]{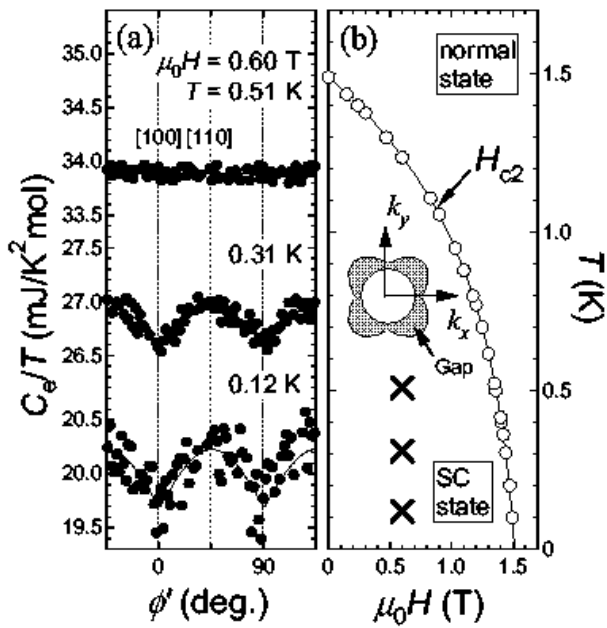}
    \end{center}
\caption{\label{fig:SRO}(a) The in-plane field-orientation dependence of the electronic specific heat divided by temperature $C_{\rm e}/T$ of \SRO\ at $\mu_0H = 0.6$ T and several temperatures. (b) The phase diagram of \SRO\ for $H \parallel [100]$, based on specific heat measurements. $H_{\rm c2}$ is the upper critical field and crosses indicate where data presented in (a) are taken.
}
\end{figure}

	The field alignment was carried out with the polar angle $\theta^{\prime}$ dependence of specific heat at several azimuthal angle $\phi^\prime$. With the choice of an appropriate field amplitude $H_{{\rm c2}\parallel ab} \gtrsim H \gg H_{{\rm c2}\parallel c}$, the SC state emerges only for the magnetic field nearly parallel to the \RO\ plane, and $C_{\rm e}/T \propto$ DOS decreases in the SC state because DOS is an increasing function of $H/H_{\rm c2}$ at low temperature, just as shown in the inset of Fig.~\ref{fig:ALHT} for aluminum. Since $H_{\rm c2}(\theta^{\prime})$ takes a maximum at $H \parallel$ \RO\ plane, $C_{\rm e}/T$ takes a minimum. Figure~\ref{fig:ALIGN}(a) shows the $\theta^\prime$ dependence of $C_{\rm e}/T$ of  \SRO\ at $T = 0.11$ K and $\mu_0H = 1.38$ T. Note that the result clearly resolves that $\mu_{0}H_{\rm c2}(\theta^\prime) > 1.38$ T only within $\mathit \Delta \theta^\prime = \pm 1.3^{\circ}$ with high angular precision. From this procedure at several azimuthal angles, we can obtain information to align $\bm{H}$ along the \RO\ plane with an accuracy better than $0.05^{\circ}$. Since the $\theta^\prime$ dependence of $C_{\rm e}/T$ is symmetric against $\theta^\prime = 90^{\circ}$, the effect of vortex pinning in the $\theta^\prime$ sweep was negligibly small for Sr$_2$RuO$_4$.

	The in-plane field alignment relative to the crystallographic axis $\phi^\prime$ was carried out, based on the known in-plane anisotropy of $H_{\rm c2}$: $H_{\rm c2\parallel [110]}>H_{\rm c2\parallel [100]}$ at low temperatures.\cite{Yaguchi,Mao} Figure~\ref{fig:ALIGN}(b) shows the $\phi^\prime$ dependence of $C_{\rm e}/T$ at $T = 0.51$ K and $\mu_0H = 1.39$ T. A sinusoidal four-fold angular variation of $C_{\rm e}(\phi)/T$ is observed. Figure~\ref{fig:ALIGN}(c) shows the field dependence of $C_{\rm e}/T$ at ${\bm H} \parallel [110]$ and ${\bm H} \parallel [100]$ at $T = 0.51$ K. In the vicinity of $H_{\rm c2}$, $C_{\rm e}/T$ takes maxima in magnetic fields parallel to the [110] directions because $H_{\rm c2\parallel [110]}>H_{\rm c2\parallel [100]}$. From these results, we can assign the directions of the crystallographic axes in the \RO\ plane with an accuracy better than $2.0^{\circ}$.

	In the mixed state, the quasiparticle (QP) energy spectrum $E_{\bm{k}}$ is affected and becomes $E^{\prime}_{\bm{k}} = E_{\bm{k}}-\delta {\it \omega}$ by the Doppler shift $\delta {\it \omega} = \hbar \bm{k} \cdot \bm{v}_s$, where $\bm{v}_s$ is the superfluid velocity around the vortices and $\hbar \bm{k}$ is the QP momentum.  In the case of $\delta {\it \omega}\geq {\mathit \Delta}(\bm{k})$, this energy shift gives rise to a finite DOS at the Fermi level.\cite{Volovik} Since the superfluid velocity $\bm{v}_s$ is perpendicular to the magnetic field $\bm{H}$, the Doppler shift $\delta {\it \omega}$ becomes zero for $\bm{k} \parallel \bm{H}$. Thus, the generation of QPs is suppressed for $\bm{H} \parallel$ gap minima directions and $C_{\rm e}/T \propto $ DOS takes minima for these directions.\cite{Vekhter} Consequently, the field-orientation dependence of the specific heat directly reveal the directions of the gap minima.

	Figure~\ref{fig:SRO}(a) shows the field-orientation dependence of the specific heat of Sr$_2$RuO$_4$. The absence of a two-fold oscillatory component in the raw data guarantees that the in-plane field alignment is accurate during the azimuthal-angle $\phi^\prime$ rotation. At $\mu_{0}H = 0.6$ T and low temperatures, shown in Fig.~\ref{fig:SRO}(b) as crosses, a nonsinusoidal four-fold angular variation approximated as $\delta C_{\rm e}(\phi)/T \propto 2|{\rm sin}2\phi^{\prime}| -1$ is observed. Since $\delta C_{\rm e}(\phi)/T$ due to the in-plane $H_{\rm c2}$ anisotropy would have an inverse phase below 1.35 T and the amplitude of the angular variation in Fig.~\ref{fig:SRO}(a) becomes smaller with an increase of temperature toward $T_{\rm c}$, we conclude that the 4-fold oscillations at a low temperature originate from the anisotropy in the SC gap.

	We have deduced the SC gap structure in the inset of Fig.~\ref{fig:SRO}(b) with a gap minimum along the [100] direction, because $C_{\rm e}(\phi)/T$ takes a minimum to the [100] direction and shows in-plane four-fold symmetry. The SC gap function is well described as ${\mathit \Delta}(\bm{k})={\mathit \Delta}_{0}({\rm sin}^2ak_x + {\rm sin}^2ak_y)^{1/2}$ with the gap minima; this SC gap structure is promising for the active SC band of Sr$_2$RuO$_4$. The details of this study are published elsewhere.\cite{Oscillation}

\section*{ACKNOWLEDGEMENTS}

The authors thank H. Yaguchi, K. Ishida, and M. Suzuki for their technical support. They also thank Y. Fujii, M. Kawakatsu, S. Takahashi, and E. Hayata for machining the calorimeter parts. They particularly acknowledge Z. Q. Mao and N. Kikugawa who grew the crystals used in our experiments. The authors also acknowledge Niki Glass Inc. for their help in constructing the rotator system and KOA Inc. for providing us with \RO\ sensors. This work was supported in part by the Grants-in-Aid for Scientific Research from JSPS and MEXT of Japan, a CREST grant from JST, and the 21COE program ``Center for Diversity and Universality in Physics'' from MEXT of Japan. One of the authors (K. D.) has been supported by a JSPS Research Fellowship for Young Scientists.

\end{document}